\documentclass[aip,rsi,amsmath,amssymb,reprint]{revtex4-2}
\usepackage{float}
\usepackage{graphicx}
\usepackage{xcolor}
\usepackage[export]{adjustbox}
\usepackage[utf8]{inputenc}
\usepackage{bm}
\usepackage[colorlinks=True,citecolor=blue,urlcolor=blue,]{hyperref}
\usepackage{indentfirst}
\usepackage{mciteplus}

\begin{document}

\title{Role of obstacle softness in the diffusive behavior of active Particles}

\author{Ankit Gupta} 
\thanks{ankitgupta@kgpian.iitkgp.ac.in}

\author{P. S. Burada }
\thanks{Corresponding author: psburada@phy.iitkgp.ac.in}
\affiliation{Department of Physics, Indian Institute of Technology Kharagpur, Kharagpur 721302, India}

\date{\today}

\begin{abstract}

We numerically investigate the diffusive behavior of active Brownian particles in a two-dimensional confined channel filled with soft obstacles, whose softness is controlled by a parameter $K$. Here, active particles are subjected to external bias $F$. Particle diffusion is influenced by entropic barriers that arise due to variations in the shape of the chosen channel geometry. We observed that the interplay between obstacle softness, entropic barriers, and external bias leads to striking transport characteristics of the active particles. For instance, with increasing $F$, the non-linear mobility exhibits non-monotonic behavior, and effective diffusion is greatly enhanced, showing multiple peaks in the presence of soft obstacles. Further, as a function of $K$ and $F$, particles exhibit various diffusive behaviors, e.g., normal diffusion - where the role of obstacles is insignificant, subdiffusion or superdiffusion - where the particles are partially trapped by the obstacles, and particles are ultimately caged by the obstacles. These findings help understand the physical situations wherein active agents diffuse in crowded environments.

\end{abstract}

\maketitle
\section{Introduction}

Microorganisms self-propel in fluid environments, and synthetic active particles (ABPs) utilize the free energy in their surroundings to convert it into sustained motion. The mechanisms driving this active motion have been extensively studied \cite{Ramaswamy,Lauga_2009, Zottl_2016, Marchetti2013, Elgeti2015,Bechinger2016,Berthier2019}. This has led to the emergence of a rapidly growing field called active matter, focusing on the physical aspects of propulsion, dynamics, and the collective behavior induced by the motility of numerous identical entities.

In natural habitats, many biological microswimmers encounter soft and solid walls, physical obstacles, and boundaries while swimming. \cite{Kaiser2012,Khatri2022} For example, microorganisms moving through porous soil columns \cite{Cortis2004}, the digestive tract\cite{2004}, 
and flow of blood cells through veins \cite{Engstler2007} face various obstacles and spatial constraints. Understanding how these microswimmers navigate and interact with their environments can provide insights into the rules that facilitate their movement in complex surroundings. 
For example, the possibility of trapping autonomously navigating microorganisms in a controlled manner offers intriguing possibilities for preventing microbial contamination and concentrating microbes near externally imposed patterned surfaces.\cite{Galajda2007}.
Theoretically, the complex heterogeneous environment is portrayed as a random Lorentz gas \cite{Hoefling2008,Zeitz2017}, wherein obstacles are randomly dispersed throughout the space, occupying a defined area fraction. The characteristics of the Lorentz gas change significantly with density. For example, with increased obstacle density, percolating clusters\cite{Mertens2012} will develop and effectively control the diffusive behavior of the particles. The clusters can arrest or imprison the particles when the system percolates at the critical obstacle density\cite{Zeitz2017,Morin2017}.
ABPs, despite being point-like, exhibit finite-size effects due to hydrodynamics, affecting their motility in crowded environments \cite{PhysRevLett.117.198001}. It is found that confinement can increase microswimmer velocity, driven by electrostatic and electrohydrodynamic effects \cite{PhysRevLett.117.198001}.

\begin{figure}[t]
\centering
\includegraphics{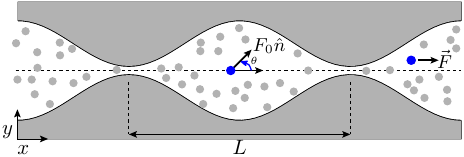}
\caption{Schematic illustration of a two-dimensional symmetric channel having periodicity $L$, described by Eq.~\ref{channel}, with a heterogeneous environment of soft obstacles. The particle self-propels with force $F_0$ and is subjected to constant force $\vec{F}$ along the $x$ direction.}
\label{fig:channel}
\end{figure}

One notable feature of these systems is the confinement that occurs due to the presence of boundaries of the chosen geometry. In undulating channels (see Fig.~\ref{fig:channel}) rather than flat channels, the irregular geometry induces spatial variations in the available phase space, leading to the emergence of entropic barriers.
These entropic barriers effectively control the diffusive behavior of particles irrespective of whether the particles are active or passive \cite{D_Reguera,Burada_2009,Borromeo2010, Borromeo2011,Gupta2023}. 
Also, by introducing changes in cross-sectional area, the undulating walls create a more complex landscape for the particles to navigate, allowing us to study how geometry can enhance or impede diffusion. This behavior is crucial for understanding how confinement affects transport in practical applications, such as microfluidic devices and porous media, where boundaries are rarely flat.

Most research has traditionally focused on hard obstacles \cite{Ai2014,Moore2023,Nayak2023,Babayekhorasani2016}, but the dynamics become even more intricate when active particles interact with deformable obstacles in confined environments, such as soft colloids or flexible biological structures \cite{Bechinger2016,Lauga_2009,D_Reguera,Dhar2020}. These deformable obstacles, more representative of real-world conditions, further alter particle trajectories, leading to deflections, directional changes, or temporary trapping.
In general, external forces, such as electric fields, magnetic fields, chemical gradients, temperature, shear flow, and acoustic waves, drive active particles through these complex landscapes. The interplay between obstacle density, the nature of obstacles, geometry confinement, and external forces results in unique transport properties that are advantageous for chemical processing, separation techniques, and catalysis. These interactions have facilitated the development of efficient filtering materials, targeted drug delivery systems, and lab-on-a-chip devices for chemical analysis and medical diagnostics \cite{Yang2017,Kaehr2009,liu2021effect,Park2008,MacDonald2003}. For example, microfluidic techniques have effectively separated living parasites from human blood \cite{Holm2011,Heddergott2012,Modica2022}. 
These experimental observations drive the development of theoretical models to comprehensively capture the behavior of systems characterized by emergent collective motion. Such models are crucial for designing and optimizing efficient lab-on-a-chip devices.

Our purpose in this article is to demonstrate that the interplay between confined geometries, deformable obstacles, and external driving forces can lead to striking transport phenomena, sometimes counterintuitive, distinct from those seen in the more familiar free space case. 
The main objective of this study is to gain insights into the transport characteristics of active Brownian particles, specifically focusing on the nonlinear mobility and effective diffusion coefficient under these conditions. 
The article is organized as follows. In Sec.~\ref{sec:model}, we introduce the Langevin model to describe the dynamics of the active Brownian particles in a two-dimensional symmetric channel. Sec.~\ref{sec:transport} is devoted to the transport characteristics of active parameters as a function of the external bias, obstacle softness, and diffusion coefficient. Also, the anomalous behavior of the effective diffusion, as a function of obstacle softness, is presented.  
Sec.~\ref{sec:discussion} and ~\ref{sec:conclusions} are devoted to the discussion and the main conclusions, respectively.

\section{MODEL}
\label{sec:model}

Consider the dynamics of an active (self-propelled) Brownian particle (ABP) suspended in a thermal bath and confined to diffuse within a two-dimensional (2D) symmetric channel, as illustrated in Fig.\ref{fig:channel}.
The particle self-propels in the direction of its orientation with a constant self-propelled velocity given by \(v_0 \hat{n}\), where \(v_0\) is the constant speed and \(\hat{n}\) is the unit vector representing the particle's orientation.
The motion of this particle is influenced by the surroundings, while within the channel resides a diverse landscape of circular soft obstacles of radius $R_0$, forming a heterogeneous environment. These obstacles do not move, effectively serving as fixed barriers in the channel.

In the over-damped regime, the Langevin equations \cite{Vicsek1995} dictating the dynamics of ABP read
\begin{align} 
\label{eq:Langeven-1}
\gamma_t \frac{d\vec{r}}{d t} &=   F_0 \hat{n} + \vec{F}_{\text{int}} + \vec{F} + \sqrt{\gamma_t k_B T} \, \vec{\xi}(t) \\
\label{eq:Langeven-2}
\gamma_r\frac{d \theta}{d t} & = \sqrt{\gamma_r k_B T} \, \chi(t)  \,.
\end{align}
Eqn.~\eqref{eq:Langeven-1} and Eqn.~\eqref{eq:Langeven-2} govern the translational and rotational motion of ABP, where $\vec{r}$ represents its position vector in 2D. The particle possesses an orientational degree of freedom characterized by the unit vector $\hat{n} = (\cos \theta, \sin \theta)$, where $\theta$ is the angle relative to the $x$ axis (channel direction).
The self-propelled force $F_0 = \gamma_t \,v_0$. Here, $k_B$ is the Boltzmann constant, and $T$ is the temperature of the surrounding medium. 
$\vec{F}_\mathrm{int}$ is the interaction force between the active particles and soft obstacles, discussed in the following sections.
$\vec{F} = F \hat{x}$ is the external force acting along the channel direction. $\gamma_t$ and $\gamma_r$ denote the translational and rotational friction coefficients, respectively.
$\vec{\xi}(t)$ and $\chi(t)$ are zero-mean Gaussian white noises which modeled the translational and rotational Brownian fluctuations due to the coupling of the particle with the surrounding medium, both obeying the fluctuation-dissipation relation $ \langle \xi_i(t)\xi_j(t') \rangle = 2\delta_{ij} \, \delta(t-t')$ for $i,j = x,y$ and $ \langle \chi(t)\chi(t') \rangle = 2\delta(t-t')$. Note that these translational and rotational noises are uncorrelated.

The half-width of the two-dimensional channel, which is symmetric and spatially periodic, describes the shape of the channel as 
\begin{equation}
    \label{channel}
	\omega(x) = a \sin \left(\frac{2\pi x}{L}\right) + b \,,
\end{equation}
where $L$ corresponds to the periodicity of the channel, $a$ determines the slope of the channel, and $b$ controls the width of the channel at the bottleneck. Specifically, we have chosen a value of $a = 1/(2\pi)$, $b = 1.02/(2\pi)$, and $L = 1$ to represent the characteristics of the channel (see Fig.\ref{fig:channel}). The minimum half-width at the bottleneck of the channel, denoted by \(\omega_{\text{min}}\), is expressed as \(\omega_{\text{min}} = b - a\), while the maximum half-width, denoted by \(\omega_{\text{max}}\), is given by \(\omega_{\text{max}} = b + a\). 
The dimensionless aspect ratio $\epsilon$ is defined as the ratio of these two widths 
$\epsilon = \omega_{\text{min}}/\omega_{\text{max}}, \,\, 0 < \epsilon \leq 1$.

Different values of \(\epsilon\) correspond to different symmetric channel shapes, e.g., \(\epsilon = 1\) represents a flat channel. The relationship between the parameters \(a\) and \(b\) in terms of \(\epsilon\) can be expressed as $b = a (1 + \epsilon) / (1 - \epsilon).$
The dynamics of the self-propelled particle at the channel walls are described as follows. The particles cannot penetrate the rigid channel walls but rotate and slide within the channel. Upon colliding with the walls, the translational velocity $\dot{\vec{r}}$ is elastically reflected, whereas the rotation angle  $\theta$ is unchanged during the collision (sliding reflecting boundary condition) \cite{Khatri2019, Ghosh2013, Reichhardt2017, Gupta2023}.  

Assuming all soft obstacles in the channel have the same radius $R_0$, the fraction of area taken by the obstacles can be estimated as $\eta = n\pi {R_0}^2/A$, where $A$ represents the area of a single cell in the channel, calculated as $2bL$. Here, $n$ represents the total number of obstacles in the cell. However, it should be noted that these obstacles may intersect or coincide with one another. Therefore, the actual area fraction \cite{book} may differ from the calculated value as $\Phi = 1- e^{-\eta}$.
In our simulations, we always choose the reduced density $\eta$ instead of the actual area fraction $\Phi$.

The particles experience elastic interactions from the obstacles $\vec{F}_\text{int}$ = $F_x \hat{e_x}$ + $F_y \hat{e_y}$. The interaction force is assumed to be of the linear spring form \cite{Khatri2020}.
\begin{equation}
\label{eq:interaction}
\vec{F}_\mathrm{int} =  K \sum_{j=1}^{n} (R_0 - r_{ij}) \hat{r_{ij}},   \ \text{for} \,\,\ r_{ij} < R_0 \,,
\end{equation}
where $K$ denotes the obstacle softness (same for all the obstacles). The summation is performed over all obstacles within the cell of the channel where the particle is located at a given moment. Furthermore, $r_{ij}$ represents the center-to-center distance between particle $i$ and obstacle $j$. When $r_{ij} < R_0$, the obstacle exerts a harmonic force on the particle. However, when \( r_{ij} \geq R_0 \), there is no interaction between the particle and the obstacle. Note that a higher $K$ value signifies more rigid obstacles, as the strength of spring force is greater (see Eq.~\ref{eq:interaction}), whereas a lower value indicates softer obstacles.

\begin{figure}[ht]
\centering
\includegraphics[scale=0.3]{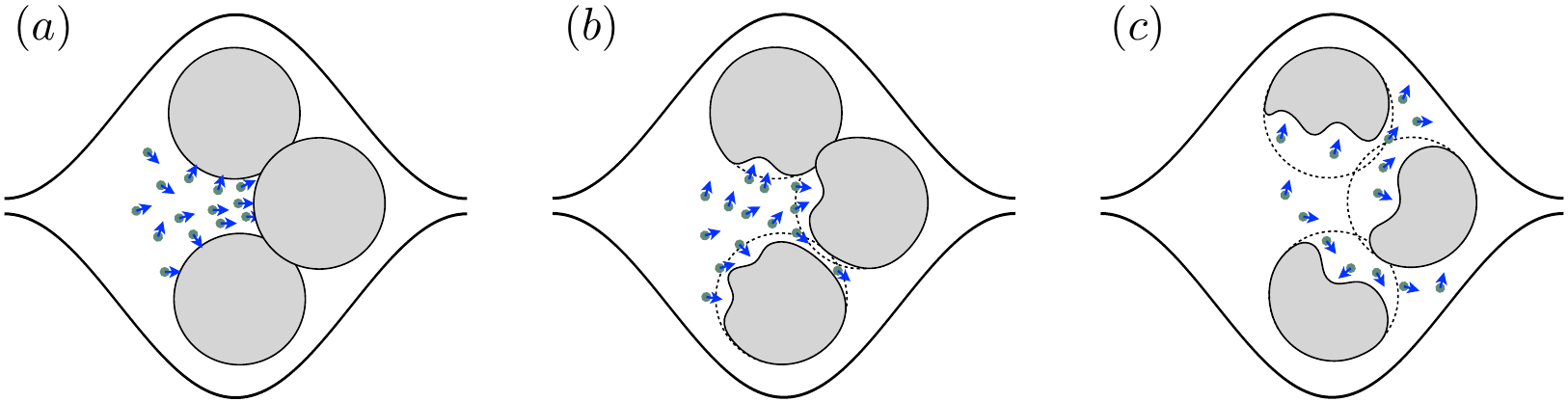}
\caption{Illustration of active particles movement in the presence of obstacles. 
Here, we have considered only three obstacles out of many for better visibility. 
(a) hard obstacles $K \to \infty$, (b) semi soft obstacles (moderate $K$), and (c) soft obstacles $K \to 0$. 
Here, $K$ denotes the obstacle softness defined in Sec. \ref{sec:model}.
}
\label{fig:sch}
\end{figure}

Figure.~\ref{fig:sch} illustrates the behavior of active particles in the presence of obstacles of various $K$. At higher $K$ values, i.e., for hard obstacles, obstacles exert a strong repulsive force, preventing particles from penetrating. Thus, particles may be trapped in the cavities formed by the obstacles depending on other model parameters. For moderate $K$ values, i.e., for semi-soft obstacles, some of the particles may squeeze through the obstacles (see Fig.~\ref{fig:sch} (b)). 
Conversely, for $K \to 0$, i.e., for very soft obstacles, particles may freely pass through the obstacles. Particles effectively do not feel the presence of obstacles (see Fig.~\ref{fig:sch} (c)). This distinction highlights how the obstacle's softness parameter $K$ regulates the interaction between active particles and obstacles in the model.

To have a dimensionless description for convenience, we scale all lengths by the periodicity of the channel $L$, translational time with characteristic translational diffusion time as $\tau_t = \gamma_t L^2/(k_B T_R)$, and rotational time with characteristic rotational diffusion time as $\tau_r = \gamma_r/(k_B T_R)$, at some reference temperature $T_R$ \cite{Burada2008}. For example, in the present work, we have taken $T_R$ as the room temperature (see Sec.~\ref{sec:discussion} for more details). However, in general, $T_R$ can be chosen depending on the model under consideration.
The dimensionless Langevin equation reads.
\begin{align} 
  \label{eq:lan-dl}
	 \frac{d\vec{r}}{d t} &=   f_0 \hat{n} + \vec{f}_{\text{int}} + \vec{f}  +  \sqrt{D_t} \, \vec{\xi}(t) \\
   \label{eq:lan-dlth}
	\frac{d\theta}{d t} & = \sqrt{D_r} \, \chi(t)  \,,
\end{align}
where, $f_0 = F_0 L/(k_B T_R)$ is the dimensionless active force, 
$\vec{f}_\mathrm{int} = \vec{F}_\mathrm{int} L/(k_B T_R)$ is the dimensionless interaction force with the dimensionless obstacle softness $k = K \,L^2/(k_B T_R)$, and $\vec{f} = f\hat{x}$ ($f = F L/(k_B T_R)$) denotes the dimensionless external bias. The dimensionless translational and rotational diffusion constants are $D_t = (k_B T /\gamma_t)(\, \tau_t/L^2) = T/T_R$ and $D_r = (k_B T/\gamma_r)\, \tau_r = T/T_R$, respectively, both of which scale with the ratio of the system temperature to the reference temperature $T_R$. This allows for a flexible exploration of different diffusion regimes, where $D_t$ and $D_r$ can be varied by a factor of 100, which is feasible and commonly used to study the system's behavior under different environmental conditions.

These dimensionless variables are used in the rest of the paper. 
Our simulations model active Brownian particles as point-sized entities with random initial conditions. 
The Langevin equations ~\eqref{eq:lan-dl}\,\&~\eqref{eq:lan-dlth} are solved using the standard stochastic Euler algorithm over $5 \times 10^2$ trajectories with boundary conditions at the channel walls as follows.
When the particle approaches the channel wall, 
the active force $\vec{F_0} = F_0 \hat{n}$ keeps pointing in the same direction. Due to the shape of the channel boundary, the particle essentially slides along the channel wall until a thermal fluctuation in the orientational vector $\hat{n}$ redirects it towards the interior of the channel \cite{Ao2014,Khatri2022}. The integration step time $\Delta$t was chosen as $10^{-8}$. In this study, we set the dimensionless radius of the obstacle $R_0 = 0.02$ for convenience. Note that obstacles packing fraction $\eta$ plays a vital role in the diffusive nature of the particles. In our earlier study, we observed that for $\eta \ge 0.4 $, particles begin to be trapped by the obstacles in the case of passive particles in a sinusoidal-shaped channel \cite{Khatri2020}. Thus, in the present study, we set $\eta = 0.4$ throughout and very the obstacle softness.

\section{Transport characteristics}
\label{sec:transport}

To characterize the transport of Brownian particles, we have computed the nonlinear mobility $\mu(f)$ and effective diffusion coefficient $D_{eff}$ defined as
\begin{equation}
    \mu(f) := \lim_{{t \to \infty}} \frac{{\langle x(t) \rangle}}{{tf}},
\end{equation}
\begin{equation}
    D_{eff} := \lim_{{t \to \infty}} \frac{{\langle x^2(t) \rangle - \langle x(t) \rangle^2}}{{2t}}.
\end{equation}
Since the external force $f$ acts on the particles along the channel direction, i.e., the $x$-axis, the above quantities are computed numerically in the same direction.

\subsection{Effect of obstacle softness}

\begin{figure}[t]
\centering
\includegraphics{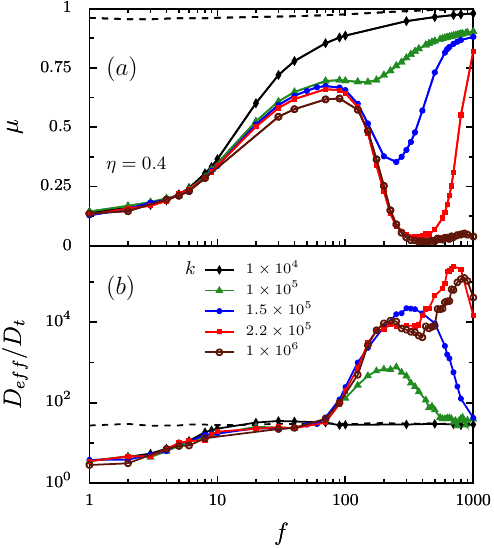}
\caption{ 
(a) The nonlinear mobility $\mu$ and (b) the effective diffusion coefficient $D_{eff}$ as a function of the external bias $f$ for various values of obstacle softness $k$. The geometry of the channel is described by Eq.~\eqref{channel}. Other parameters in the system are set as follows: $a = 1/2\pi$, $b = 1.02/2\pi$, $D_t = 1$, $D_r = 1$, $f_0 = 1$, and $ R_0 = 0.02$. Dashed lines represent the case of a flat channel ($\epsilon = 1$) for $k = 1 \times 10^4$.}
\label{fig:mu_deff_f}
\end{figure}

As mentioned earlier, obstacles are placed randomly in the channel. The particles exhibit diffusive transport under the influence of self-propelled force, random force due to noise, and the applied external force. We began by altering the obstacle's softness, controlled by the parameter $k$ (see Eq.~\eqref{eq:interaction}). Fig.~\ref{fig:mu_deff_f} depicts the nonlinear mobility $\mu$ and the scaled effective diffusion coefficient $D_{eff}$ as a function of the external bias $f$ for various values of $k$. It is evident that $k$ greatly influences $\mu$ and $D_{eff}$. 
Note that for a given dimensionless equation Eq.~\eqref{eq:lan-dl} \& \eqref{eq:lan-dlth}, we find that up to $k\sim 10^4$ obstacles are extremely soft and have little impact on the particle diffusion. In the case of a flat channel ($\epsilon = 1$) with extremely soft obstacles, $\mu$ slowly increases with $f$ and finally reaches the value one in the limit $f \to \infty$. $D_{eff}$ also shows a similar trend and approaches a finite value in the limit $f \to \infty$. See the dashed lines in Fig.~\ref{fig:mu_deff_f}. 
In the $f \to 0$ limit, $\mu < 1$ is due to the presence of the obstacles, even though they are soft. 
However, for $\epsilon \ll 1$, as the case in the current study, entropic barriers arise due to the geometrical confinement \cite{Burada_Chemphy}, as a result, $\mu$ having the lower values and $D_{eff}$ is enhanced compared to the flat channel (see Fig.~\ref{fig:mu_deff_f}).

In the case of modulated channels with extremely soft obstacles (see Fig.~\ref{fig:sch} (c)), $\mu$ increases monotonically with $f$, similar to the case of passive particles \cite{Reguera2006, Borromeo2011}. 
However, $D_{eff}$ increases monotonically with $f$ and finally approaches the value 
corresponding to the flat channel at higher $f$ strengths. 
However, it is predictable that active particles exhibit higher effective diffusion than passive particles due to their rotational degrees of freedom. The rotational diffusion causes a randomization of the propulsion direction, resulting in a random walk with a step length proportional to the product of the propulsion velocity $v_0$ and the rotational diffusion time $\tau_r$. This process significantly enhances the effective diffusion coefficient beyond the translational diffusivity $D_t$ \cite{Bechinger2016}. 

Interestingly, as $k$ increases, i.e., as the softness of the obstacles decreases (Fig. \ref{fig:sch} (b)), $\mu$ exhibits a non-monotonic behavior, and the corresponding $D_{eff}$ is greatly enhanced (see Fig.~\ref{fig:mu_deff_f}(b)). It is because as $k$ increases, the particle's movement is hindered by the obstacles. For smaller values of $f$, particles can easily escape from the obstacles due to the dominance of their intrinsic active nature and thermal fluctuations, and $\mu$ increases monotonically. $D_{eff}$ also shows a similar trend. However, as $f$ increases, particles are stuck in the soft cavities formed by the obstacles. Thus, $\mu$ decreases depending on the strength of $k$. Surprisingly, $D_{eff}$ is greatly enhanced and shows multiple peaks for moderate $f$ values. It is because the diffusive nature of the active particles changes with the obstacle softness $k$ and external bias $f$. The diffusion of particles may not be normal. The hindering motion of particles in the presence of obstacles may differ from normal diffusion, i.e., sub or super-diffusive behavior. The latter is investigated in subsection~\ref{subsec:Anamalous}, see Fig.~\ref{fig:alpha}(b). Further increasing $f$ helps the particle to escape from the obstacles, and as a result, $\mu$ increases steeply, and $D_{eff}$ tends to show normal behavior, i.e., exhibiting behavior similar to that observed in the absence of obstacles in the channel.

Further, an increase in $k$ makes obstacles hard (see Fig. \ref{fig:sch} (a)). As a result, obstacles impede particle motion. Particles are trapped in the hard cavities formed by the obstacles. If the applied force on the particles is weak, then they can escape from the cavities due to their active nature and thermal fluctuations. However, when $f$ is very strong, where activity and thermal fluctuations are suppressed, particles are stuck in the cavities, resulting in zero mobility. Correspondingly, $D_{eff}$ slows down (see Fig.~\ref{fig:mu_deff_f}(b)). A similar trend is evident in the case of passive particles in the presence of solid obstacles \cite{Khatri2020}. 

\subsection{Influence of translational and rotational diffusion constants}

\begin{figure}[ht]
\centering
\includegraphics{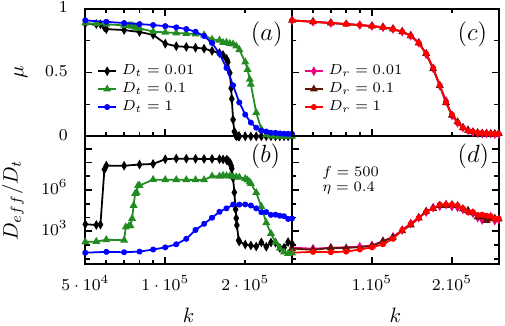}
\caption{
(a) The nonlinear mobility $\mu$ and (b) the effective diffusion coefficient $D_{eff}$ as a function of obstacle softness $k$ for various strengths of translational diffusion constant $D_t$. (c) The nonlinear mobility $\mu$ and (d) the effective diffusion coefficient $D_{eff}$ as a function of $k$ for various strengths of rotational diffusion constant $D_r$. Here, we set $\eta = 0.4$ and $f = 500$. Other parameters in the system are set as follows: $a = 1/2\pi$, $b = 1.02/2\pi$, $f_0 = 1$, and $ R_0 = 0.02$.}
\label{fig:d0_dr}
\end{figure}

\begin{figure}[!ht]
\centering
\includegraphics{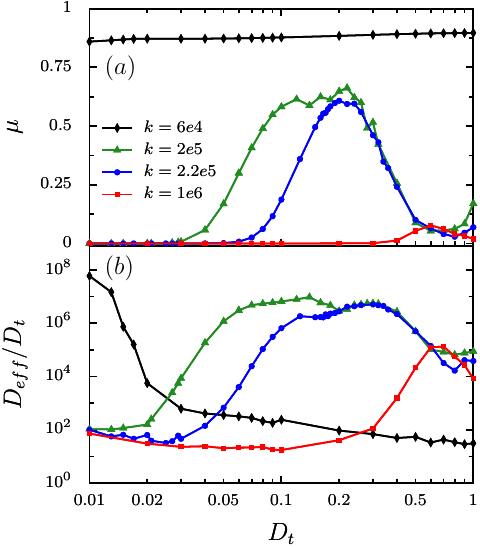}
\caption{$(a)$ The nonlinear mobility $\mu$ and $(b)$ the effective diffusion coefficient $D_{eff}$ as a function of the translational diffusion constant $D_t$ for various values of obstacle softness $k$. Here, we set $f =500$, $D_r =1$ and $\eta = 0.4$.
The other set parameters are $a$ = 1/2$\pi$, $b$ = 1.02/2$\pi$, $f_0 = 1$, and $R_0$ = 0.02.}
\label{fig:mu_deff_Dt}
\end{figure}

Figure~\ref{fig:d0_dr} shows the behavior of $\mu$ and $D_{eff}$ as a function of $k$ for different strengths of translational ($D_t$) and rotational diffusion ($D_r$) constants. 
Remarkably, from Fig.~\ref{fig:mu_deff_Dt}, we can define a critical $k$ value $(k_c \approx 2 \times 10^5)$ in the deterministic limit and at a moderate force value $f = 500$, where $\mu$ and $D_{eff}$ show a sharp decay. 
For $k < k_c$, i.e., when the obstacles are relatively soft (Fig. \ref{fig:sch} (c)), $\mu$ decays slowly with increasing $k$. However, for $k > k_c$, obstacles become hard (Fig. \ref{fig:sch} (a)), and as a result, $\mu$ decreases sharply with increasing $k$ (see Fig.~\ref{fig:d0_dr}(a)). Note that this trend is independent of $D_r$ (see Fig.~\ref{fig:d0_dr}(c)); however, $D_t$ shows some impact. On the other hand, in the $k \to 0$ limit, particles do not feel the presence of obstacles and exhibit normal diffusion (see Fig.~\ref{fig:d0_dr}(b) and (d)). With increasing $k$, $D_{eff}$ increases slowly with $k$. However, when $k$ values of significance, $D_{eff}$ sharply increase show either subdiffusive or superdiffusive behavior (see Fig.~\ref{fig:alpha}) for $k < k_c$. In this range, the obstacles are relatively soft. For $k > k_c$, obstacles are hard, and particles are trapped by the obstacles. As a result, $D_{eff}$ is constant and $\mu$ tends to zero. As mentioned before, the rotational diffusion constant $D_r$ does not influence the behavior of $D_{eff}$, but $D_t$ does.

Figure~\ref{fig:mu_deff_Dt} depicts the behavior of $\mu$ and $D_{eff}$ as a function of the translational diffusion constant $D_t$ for various values of $k$. For the scenario of very soft obstacles, a higher value of non-linear mobility is evident \cite{Ao2015}. The corresponding $D_{eff}$ exponentially decays. This relationship is particularly interesting as it mirrors the behavior of active particles diffusing in a $2D$ channel without obstacles. The motion of the particles in this situation exhibits a straightforward and predictable trend, emphasizing the role of obstacle softness in shaping $\mu$ and $D_{eff}$. In the case of moderate $k$, $\mu$ initially rises, reaching a peak at a moderate $D_t$ value, and subsequently decreases, forming a distinctive bell curve (Fig.~\ref{fig:mu_deff_Dt}(a)). The corresponding $D_{eff}$ shows a similar trend (Fig.~\ref{fig:mu_deff_Dt}(b)). This nuanced pattern highlights the relationship between $k$ and $D_t$, suggesting an optimal range of $D_t$ for maximizing $\mu$. For very high $k$ values, particles are mostly trapped by the obstacles, resulting in $\mu \to 0$, but finite $D_{eff}$ due to the contribution from the rotational diffusion constant $D_r$ (Fig.~\ref{fig:mu_deff_Dt}(b)). Therefore, there is an optimal value of $D_t$ at which $\mu$ takes its maximal value, and the corresponding $D_{eff}$ is also high.

\begin{figure}[ht]
\centering
\includegraphics[width=\linewidth]{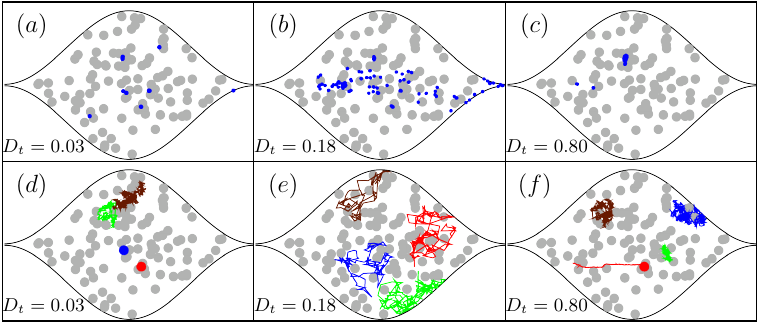}
\caption{Steady state behavior of active Brownian particles for various values of the translational diffusion constant $D_t$ with obstacle softness $k = 2.2 \times 10^5$, $f=500$, and $D_r =1$. Here, the particle's position is mapped into a single cell of the $2D$ channel filled with the heterogeneous distribution of obstacles. The other set parameters are $a$ = 1/2$\pi$, $b$ = 1.02/2$\pi$, $\eta =0.4$, $f_0 = 1$, and $R_0$ = 0.02.}
\label{fig:steady}
\end{figure}

Figure ~\ref{fig:steady} depicts the steady state behavior of the particles in the presence of obstacles for different $D_t$ values. Fig.~\ref{fig:steady}(a - c) shows the position of the particles in the steady state while Fig.~\ref{fig:steady}(d - f) shows the trajectories of a few particles in the steady state. 
Given the periodic nature of the channel and the heterogeneous distribution of obstacles within a cell, the particle's position has been mapped into a single cell for clarity. Figure~\ref{fig:steady} shows that obstacles disrupt the trajectories, forcing the particles to either slow down, halt, or change direction, thus effectively reducing their overall mobility and effective diffusion. Consisting with the behavior observed in Fig.~\ref{fig:mu_deff_Dt}, for smaller values of $D_t$, particles show less diffusivity (see Fig.~\ref{fig:steady}(a)\&(d)). As $D_t$ increases, particles spread around and lead to increased randomness in particle trajectories. As a result, mobility and effective diffusion increase (see Fig.~\ref{fig:steady}(b,c,e,f) \& Fig.~\ref{fig:mu_deff_Dt}(a,b)). At higher values of $D_t$, the particles make frequent collisions with the obstacles and may get stuck in the cavities formed by the obstacles (see Fig.~\ref{fig:steady}(c)\&(f)) if the obstacles are hard enough. Otherwise, they may escape from the cavities due to their intrinsic activity and thermal fluctuations. As a result, particles show finite $\mu$ and $D_{eff}$ as it is evident from Fig.~\ref{fig:mu_deff_f}, ~\ref{fig:d0_dr}, \& ~\ref{fig:mu_deff_Dt}.

\subsection{Anomalous diffusion of active particles - impact of obstacles}
\label{subsec:Anamalous}

While studying effective diffusion, we focus on the variance in the particle position along the channel direction over time, denoted by $\langle x^2(t) \rangle = \langle x(t)^2 \rangle - \langle x(t) \rangle^2$. This variance follows a power law relationship with time, expressed as $\sim t^{\alpha}$, where $\alpha$ determines the nature of the diffusion process. Normal diffusion, as described by a linear function over time ($\alpha = 1$), is a typical scenario \cite{Bouchaud1990}. However, deviations from this linear behavior at longer times indicate anomalous diffusion \cite{Bouchaud1990, Metzler2000} encompasses various behaviors: subdiffusion occurs when $0 < \alpha < 1$, while superdiffusion is observed for $1 < \alpha < 2$. Additionally, $\alpha = 0$ corresponds to the trapped state. One can numerically compute the local exponent $\alpha(t)$ as \cite{Zeitz2017}
\begin{align}
\alpha(t) = \frac{d \log[\langle x(t)^2 \rangle - \langle x(t) \rangle^2]}{d \log t}\,,
\label{eq:alpha}
\end{align}
and the steady-state values of the exponent can be obtained as $ \alpha = \lim_{t \to \infty} \alpha(t) $. As time approaches infinity, these steady-state values reflect the long-term behavior of the system, offering insights into its overall dynamics.

\begin{figure}[ht]
\centering
\includegraphics[scale=0.5]{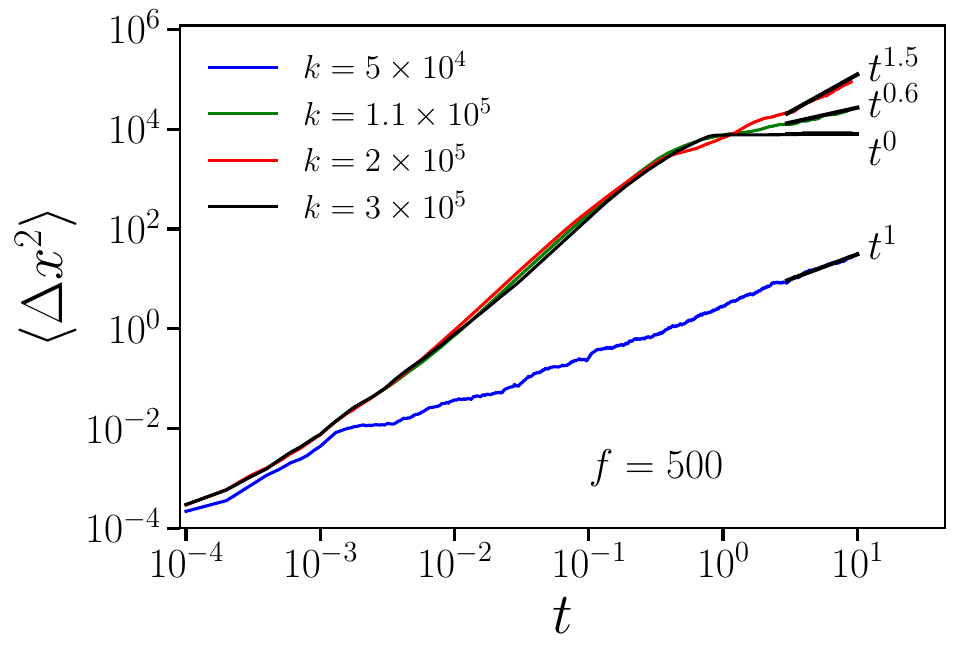}
\caption{Variance $\langle \Delta x^2(t)\rangle$ as a function of time $t$ for different obstacle softness $k$. Here, we set $D_t =1$, $D_r =1$, and $\eta = 0.4$. Additional parameters in the system are set as follows: $a = 1/2\pi$, $b = 1.02/2\pi$, $f_0 = 1$, and $ R_0 = 0.02$.}
\label{fig:MSD}
\end{figure}

Fig.~\ref{fig:MSD} shows the variance $\langle x^2(t) \rangle$ as a function of time for different obstacle softness $k$ and for a fixed strength of external bias. 
Interestingly, $\langle x^2(t) \rangle$ strongly depends on $k$. It shows that particles exhibit normal diffusion ($\alpha=1$) in the presence of soft obstacles, as expected. However, interestingly, as $k$ increases, i.e., when the obstacles are semi-soft, $\langle x^2(t) \rangle$ shows anomalous diffusion (Fig.~\ref{fig:MSD}). In the case of hard obstacles, particles are trapped ($\alpha=0$) in the cavities formed by the obstacles. 

\begin{figure}[ht]
\centering
\includegraphics{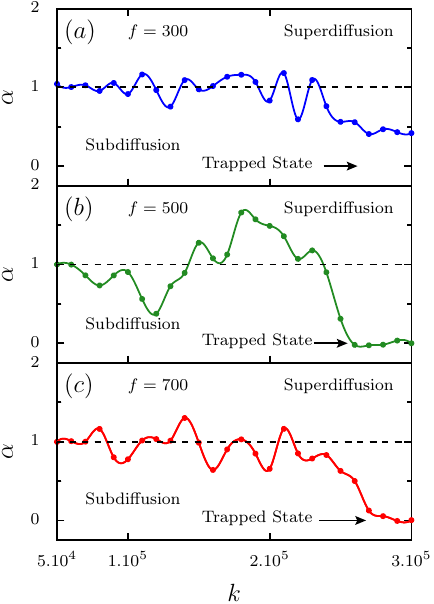}
\caption{The exponent $\alpha$ as a function of the obstacle softness $k$ for various values of the external bias $f$ with $D_t =1$, $D_r =1$ and $\eta = 0.4$. Additional parameters in the system are set as follows: $a = 1/2\pi$, $b = 1.02/2\pi$, $f_0 = 1$, and $ R_0 = 0.02$.}
\label{fig:alpha}
\end{figure}

Figure.~\ref{fig:alpha} shows the local exponent $\alpha$ as a function of $k$ for various strengths of the external bias $f$. For $k \ll k_c$, i.e., when the obstacles are relatively soft (see Fig. \ref{fig:sch} (c)), particles mostly show normal diffusion with $\alpha$ value close to one. With increasing $k$, particles show subdiffusion or superdiffusion depending on the strength of $f$. Interestingly, particles exhibit normal diffusion in the regime of weak and strong bias $f$ (see Fig.~\ref{fig:alpha}(a) \& (c)), particularly when the obstacles have some softness. For moderate strength of $f$, e.g., $f = 500$, due to the interplay between $k$ and $f$, particles show subdiffusive or superdiffusive behavior with $\alpha$ varying between $\sim[0.5:2]$. See Fig.~\ref{fig:alpha}(b). Obstacles can lead to subdiffusive behavior, where the variance grows slower than linearly with time due to trapping and frequent collisions. However, the observed superdiffusive behavior, where the variance grows faster than linearly with time, is due to long, directed runs interrupted by particle collisions with the obstacles. At higher $k$ values, i.e., when the obstacles are hard (see Fig. \ref{fig:sch} (a)), particles get trapped in the cavities formed by the obstacles as depicted in Fig.~\ref{fig:steady}(d).

\section{Discussion}
\label{sec:discussion}

From the above observations, it is clear that the transport characteristics, including the nonlinear mobility and effective diffusion coefficient, are greatly affected by both the obstacle softness $k$ and the external bias $f$ within the confined geometry. Furthermore, our observations show that the particles exhibit various diffusion behaviors, including normal diffusion, anomalous diffusion, and trapped states. The aforementioned finding can be studied experimentally for active particles diffusing in a symmetric channel. This channel \ref{channel} can be created by microprinting on a substrate \cite{Yang2017}. To provide an estimate in real units, which is highly useful for experimentalists, for particles moving in a symmetric channel with aspect ratio $\epsilon = 0.1$ and period length $L \sim 10 \, \mu m$ filled with water at room temperature ($T_R \sim 300$ K) having viscosity $\eta \sim 2 \times 10^{-3} mg/s$ and characteristic diffusion time  $\tau \sim 50 \,s$.
Particles moving with $v_0 \sim 0.2 \,\mu m/s$, in the presence of obstacles of size $\sim 0.2 \mu m$, exhibit normal diffusion under weak and strong bias regimes ($F \sim 0.12 \ pN $ and $F \sim 0.3 \ pN $, respectively). However, particles demonstrate subdiffusive and superdiffusive behavior for intermediate bias values ($F \sim 0.2 \ pN $). Furthermore, obstacles exhibit relative softness at the critical $K \sim 8 \ pN/\mu m $, resulting in subdiffusive or superdiffusive particle behavior \cite{Morin2017}. Conversely, obstacles become hard above this $K$ value, leading to particle trapping. These results will motivate the experimentalists to design laboratory-on-a-chip devices and artificial channels for controlled drug release, particle transport, and particle separation. \cite{Bechinger2016,Volpe2011,Yang2017, D_Reguera,Kaehr2009,liu2021effect,Park2008,MacDonald2003}.

\section{CONCLUSION}
\label{sec:conclusions}

In conclusion, our study demonstrates that the softness of obstacles, controlled by the parameter $k$, significantly influences the nonlinear mobility $\mu$ and the effective diffusion coefficient $D_{eff}$ of particles undergoing diffusive transport. For soft obstacles ($k \to 0$), particles experience minimal impact from the obstacles, leading to a monotonic increase in $\mu$ and a bell-shaped $D_{eff}$ curve with external bias $f$. We have observed that semi-soft obstacles form soft cavities, where $\mu$ exhibits non-monotonic behavior, and $D_{eff}$ is greatly enhanced, displaying multiple peaks for moderate $f$ values. It is a clear signature of the semi-soft obstacles. Additionally, in the limit $f \to \infty$, $\mu$ shows a rapidly increasing behavior. Also, for the optimal values of the translational diffusion constant $D_t$, $\mu$ gets maximized. Particles are significantly hindered by hard obstacles, leading to zero mobility and weak diffusion under strong external bias. Furthermore, it is revealed that, as a function of $k$ and $f$, particles exhibit various diffusive behaviors, including normal diffusion, anomalous diffusion, and trapped states. 
The important point to be noted is how the soft obstacles control the diffusive nature of the particles. 
For example, for moderate strengths of $f$, particles exhibit subdiffusive or superdiffusive behaviors in the presence of soft or semi-soft obstacles. This clearly impacts the softness of the obstacles. 
These findings are relevant for a broad spectrum of physical systems wherein active agents are self-propelled through diverse environments, e.g., the movement of tracer particles in a cell, laboratory-on-a-chip devices, and artificial channels. Further, this study may be useful in controlled drug release, particle transport in crowded environments, and particle separation \cite{Bechinger2016,Volpe2011,Yang2017, D_Reguera,Kaehr2009,liu2021effect,Park2008,MacDonald2003}.

\section{ACKNOWLEDGMENT}

The authors acknowledge the financial support by SERB, DST, the Government of India, and Project-wide CRG/2023/006186. The authors acknowledge the National Supercomputing Mission (NSM), Government of India, for the computing resources of PARAM Shakti at IIT Kharagpur.

\section{Data availability}

The data that support the findings of this study are available from the corresponding author upon reasonable request.




%

\end{document}